\documentclass[preprint2]{aastex}
\usepackage{graphicx}
\usepackage{amsmath}

\newcommand{\kms}{{\, \rm km~s}^{-1}}
\newcommand{\ha}{H$\alpha$}
\newcommand{\hb}{H$\beta$}
\newcommand{\nd}{[NII] }

\shortauthors{Letawe et al.}

\begin{document}
\title{Understanding the relations between QSOs and their host galaxies from combined HST imaging and VLT spectroscopy\altaffilmark{1}.}

\author{Y. Letawe\altaffilmark{2}, P. Magain\altaffilmark{2}, G. Letawe\altaffilmark{2}, F. Courbin \altaffilmark{3} and D. Hutsem\'ekers\altaffilmark{2,4}}

 \email{yletawe@ulg.ac.be}
\altaffiltext{1}{based on observations made with the Nasa/ESA Hubble Space Telescope (cycle 13 proposal \#10238), and with ANTU/UT1 at ESO-Paranal observatory in Chile (programs 65.P-0361(A) and 66.B-0139(A)).}
\altaffiltext{2}{Institut d'Astrophysique et de G\' eophysique, Universit\' e de
  Li\`ege, All\'ee du 6 Ao\^ut, 17, Sart Tilman (Bat. B5C), Li\`ege, Belgium.}
\altaffiltext{3}{Ecole Polytechnique F\'ed\'erale de Lausanne (EPFL), Laboratoire d'Astrophysique, Observatoire, CH-1290 Sauverny, Switzerland.}
\altaffiltext{4}{Senior research associate FNRS.}
\begin{abstract}

The host galaxies of six nearby QSOs are studied on the basis of high resolution HST optical images and spatially resolved VLT slit spectra. 
The gas ionization and velocity are mapped as a function of the distance to the central QSO. In the majority of the cases, the QSO significantly contributes to the gas ionization in its whole host galaxy, and sometimes even outside.

Reflection or scattering of the QSO \ha\ line from remote regions of the galaxy is detected in several instances. The line shifts show that, in all cases, the matter responsible for the light reflection moves away from the QSO, likely accelerated by its radiation pressure.

The two faintest QSOs reside in spirals, with some signs of a past gravitational perturbation.  One of the intermediate luminosity QSOs resides in a massive elliptical containing gas ionized (and probably pushed away) by the QSO radiation.  The other medium-power object is found in a spiral galaxy displaying complex velocity structure, with the central QSO moving with respect to the bulge, probably as a result of a galactic collision.  The two most powerful objects are involved in violent gravitational interactions and one of them has no detected host.

These results suggest that (1) large-scale phenomena, such as galactic collisions, are closely related to the triggering and the feeding of the QSO and (2) once ignited, the QSO has significant influence on its large-scale neighborhood (often the whole host and sometimes further away).

\end{abstract}


\section{Introduction}

The study of QSOs and their host galaxies has become a widespread subject of investigation in astrophysics for the past 10 years. Some recent developments show that statistical analyses over samples of QSOs may lead to fruitful conclusions about the morphology, the gaseous and stellar content of the host galaxies, and about the mass of the central black hole, the QSO ignition and fueling,etc.  However, no clear scheme has emerged yet to account for the diversity of QSO and host galaxies properties and the debate on their interrelations is still widely open.

Previous studies of QSO host galaxies were carried out using either spectroscopic data \citep{Hughes,Baldwin,Letawe,VdB,Kewley} or high resolution images \citep{Bahcall97,Dunlop,Floyd}. None of them used both spectroscopy and high resolution imaging.

This kind of study was carried out for the bright QSO HE 0450--2958 \citep{Mag2} showing a strongly distorted galaxy, about 7 kpc away from the QSO.  Previous studies \citep{Boyce96,Canalizo2001}, assumed that it probably experienced a collision with the QSO host galaxy.  A careful processing of the HST images allowed to separate the point source from the extended objects, with the surprising result that no host galaxy could be detected around the QSO.  Only a compact `blob' next to the QSO could be found, whose spectrum was shown to consist in a cloud of gas ionized by the QSO radiation, with no trace of a stellar component. The nature of HE 0450--2958 is still controversial: is there no host galaxy at all, or could it be faint enough to escape detection? Which formation mechanism could account for such a special configuration? Whatever the answer, the study of HE 0450--2958 is a beautiful example of the nice complementarity between spectroscopy and imaging. 

The aim of the present paper is to extend this kind of analysis to the six Type 1 QSOs for which we have both high resolution images obtained with the Hubble Space Telescope (HST) and slit spectra using FORS1 on the ESO Very Large Telescope (VLT) \citep{Letawe}. The six QSOs are: HE 0306--3301, HE 0354--5500, HE 0450--2958, HE 1434--1600, HE 1503+0228 and HE 2345--2906.

Section 2 gives an overview of the observational data and of the main characteristics of the reduction process. In Sect.\  3, the post-processing of the HST images is explained, while Sect.\  4 describes the extraction of the spatially resolved host spectra.  In Sect.\  5, the methods used to analyze simultaneously the images and the spectra and to infer the host properties are explained. Section 6 presents the results on individual objects, while Sect.\  7 provides a general discussion and some conclusions on the QSO host properties. Throughout the paper, we adopt the following cosmology: $H_0=65\kms $Mpc$^{-1}$, $\Omega_m=0.3$ and $\Omega_\lambda = 0.7$.

\section{Observations and reduction}

\subsection{The sample}
The six QSOs form a subsample of the 20 bright low redshift QSOs ($z < 0.35$, $M_V <-23$) from the Hamburg-ESO Survey analysed by \citet{Letawe} for which we also have high resolution images obtained with HST. The subsample was selected on the basis of their spectra with preference to those revealing large amount of gas and interesting features, which introduces unavoidable biases. Moreover, It is obviously too small to attempt any statistical analysis. Thus, the present analysis can be seen as (1) a demonstration of the interest of combining high resolution images and spatially resolved spectra to study the QSO hosts and (2) the focus on interesting cases, which can shed light on the relations between the QSOs and their hosts galaxies or, more generally, their environment.

\subsection{Images}
The images have been obtained in September-October 2004 and April 2005 with the Advanced Camera for Surveys/High Resolution Channel onboard HST with the broad V-band filter F$606$W. The ACS/HRC detector consists of a $1024 \times 1024$ pixels CCD,  with $\sim0.028 \arcsec \times 0.025\arcsec $ per pixel, covering a $29\arcsec \times 25\arcsec $ field of view.

\begin{figure}
\centering
\includegraphics[width=11 cm]{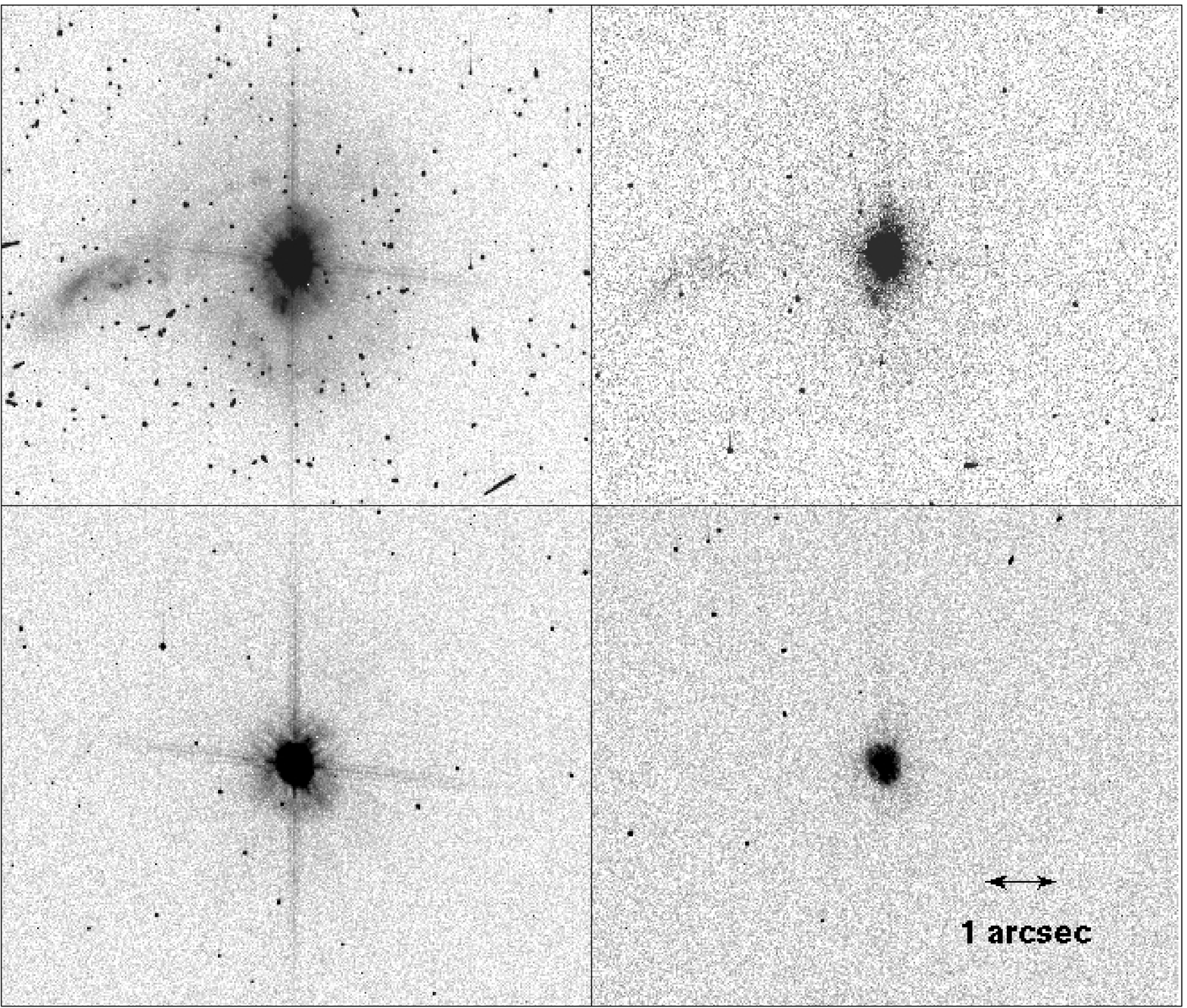}
\caption{Flatfielded observations of HE0354-5500 and a PSF star with HST/ACS. Top left: QSO long exposure. Top right: QSO short exposure. Bottom left: PSF star long exposure. Bottom right: PSF star short exposure.}
\label{cl}
\end{figure}

For each QSO, six exposures, taken during one HST orbit, are available: three short unsaturated ones, in order to be able to study the host properties very close to the central QSO, and three long ones, saturated around the center, but providing a good S/N in the lower surface brightness regions. Small offsets are applied to ease cosmic rays and bad pixels removal (5 pixels in each direction between successive exposures). Moreover, a PSF star is observed after each QSO (Fig. \ref{cl}).  Table \ref{sampl} summarizes the main QSO characteristics and the characteristics of the HST observations.

\begin{table*}
\centering 
\caption[]{Observational characteristics of the sample (redshift, absolute V-band magnitude of the QSO $M_V$ from the VLT observations \citep{Letawe}, short and long exposure times and position angle (PA) of the observation, counted positive East of North.}
\begin{tabular}{lccccc}
\hline
Name &Redshift&$M_V$(QSO) &Short E.T.&Long E.T.&PA \\ 
     &     &   &  (sec)&(sec)&(\degr) \\
\hline\hline
HE2345-2906 & $0.223$& $-22.97$ &  $40$ & $340$&$177.8$\\
PSF & & &$10$ & $30$&\\
HE1503+0228  & $0.135$ & $-23.43$ & $40$ & $330$&$36.8$  \\
PSF & & &$10$ & $40$&\\
HE1434-1600  & $0.144$& $-23.82$ & $30$ & $330$&$288$ \\
PSF & & &$5$ & $40$&\\
HE0306-3301 & $0.247$& $-24.81$ & $30$ & $330$&$110$\\
PSF & & &$15$ & $45$&\\
HE0354-5500 & $0.267$& $-25.45$ & $40$ & $400$&$258$\\
PSF & & &$5$ & $40$&\\
HE0450-2958  & $0.285$& $-25.82$ & $30$ & $330$ &$105.6$ \\
PSF & & &$6$ & $45$& \\
\hline
\end{tabular}
\label{sampl}
\end{table*}
\placefigure{cl}

The data are provided under different forms, along with auxiliary files:\\
-- the observation itself (available in raw, flatfielded, or drizzled form)\\
-- a flag image indicating hot, warm or bad pixels, and cosmic rays hits\\
-- a standard deviation image.

The drizzled data are the individual raw images corrected from optical distorsions inherent to the ACS instrument and combined into a single image. This correction uses interpolations. We avoid their use in order to lower the number of manipulations, in particular interpolations which introduce correlated noise in adjacent pixels. We rather correct for distorsion at the very end of the process, on the subsampled deconvolved images. Our data processing method (MCS, see below) allows both to simultaneously deconvolve several dithered images of the same field and to oversample the data by a factor of 2. It thus performs its own optimal drizzling. 
The image reduction itself is performed on the flatfielded data using the Pyraf software.
A standard deviation map is created using, for every pixel $i$ of an observation, the formula
\begin{equation}
\sigma_{i}= \sqrt{\frac{A_i}{g} + \big(\frac{r}{g}\big)^2},
\end{equation}
where $A_i$ is the signal in pixel $i$, $g$ the CCD gain, and $ r$ the background (readout) noise. Then we create an inverse sigma image. This allows to give stricly nil weight to bad or saturated pixels and cosmic ray hits, simply by attributing them a zero value on the inverse sigma image. This corrected inverse ``noise-image" is used for the MCS processing.

\subsection{Spectra}

The spectroscopic data were obtained with the ESO VLT/UT1 in April and November 2000 with FORS1 in Multi Object Spectroscopy (MOS) mode, which consists in a set of $1''$-slits with a fixed length of $19''$. Every QSO was observed with $3$ different filters, covering a total restframe wavelength band of $5000$\AA, from $3500$\AA\ to $8500$\AA.  One slit was centered on the QSO and the other ones on PSF stars and surrounding objects. 

The reduction and deconvolution methods are presented in detail in \citet{Letawe}, resulting in spatially deconvolved spectra with the QSO spectrum separated from the spectrum of its host galaxy.  This spatial separation of the point-like component from the diffuse one relies on the hypothesis that the latter contains no significant structure narrower than the fixed and finite resolution of the deconvolved spectrum. In order to regularize the solution of this inverse problem, one has to apply a smoothing constraint on the diffuse component.

Such inevitable smoothing decreases the spatial resolution of the host galaxy spectrum, especially in cases where the host S/N is low (the higher the S/N, the better the spatial resolution of the host spectrum).  To avoid this variable-resolution problem, we proceed in a slightly different manner than in \citet{Letawe}.  Instead of relying on the host spectrum as given by the deconvolution process, we start from the original (reduced) spectra and {\em subtract} the two-dimensional QSO-alone spectrum given by deconvolution.  Thus, we get a 2D host galaxy spectrum with a spatial resolution constant along the slit and equal to that of the original data.

\section{Image processing}
The HST/ACS images are processed with the MCS deconvolution algorithm \citep{Mag}, which is particularly well suited for separating point sources from diffuse ones and, as such, very well adapted to the study of QSO host galaxies.  Indeed, as demonstrated earlier in the case of HE 0450--2958 \citep{Mag2}, the extended wings of the HST PSF may cause the light from the bright QSO to significantly contaminate the image of the nearby souces.  Important components can be hidden, such as ``the blob" in the aforementioned case.

The principle of the MCS method is to produce images with a resolution and sampling fixed by the user, provided that they conform to the sampling theorem.  In our case, the deconvolved image has a pixel size two times smaller than the original HST images and a PSF of Gaussian shape, with a Full Width at Half Maximum (FWHM) of two final pixels.  The point sources are explicitely separated from the diffuse components, which allows to get an image of the host galaxy uncontaminated by the QSO (see Fig. \ref{deconv}).

\subsection{PSF construction}
A crucial point in the deconvolution process is the PSF construction: the more accurate the PSF, the better the deconvolution. In order to construct the best possible PSF, we use the method explained in \citet{Let06}. Here are its main steps:

(1) With the TinyTim software\footnote{Krist J.,\\http://www.stsci.edu/software/tinytim/tinytim.html}, we create a synthetic PSF corresponding to the actual observing conditions. Let us call it $TT$. We then compute the kernel which allows to transform the $TT$ image into a gaussian $r$ of 2 pixels FWHM.  Let us call this kernel $TinyDec$. It obeys the equation:

$TT(x)=TinyDec(x) \ast r(x)$, 

where $x$ is a 2D vector representing the position in the image and $\ast$ stands for the convolution operator. 

(2) $TT$ is a model of the actual PSF.  However, it is not accurate enough to provide a trustworthy separation of the QSO from its host galaxy.  We improve it by comparing $TinyDec(x) \ast r(x)$ to the observation $t(x)$ of a point source (one of the reduced PSF star observations). In order to improve the fit, we add a numerical component $F(x)$ to the model $TinyDec(x)$ until the observed PSF star image is reproduced with satisfying accuracy:
\begin{equation}
t(x)=(TinyDec(x)+F(x))\ast r(x).
\end{equation}
Among all the PSF stars, we choose the one which is best suited for a given QSO observation. As the HST PSF depends on a number of parameters (such as observed target spectral distribution, image position in the focal plane, position of the satellite on its earth orbit,\ldots) we test a variety of QSO/PSF combinations and choose the one which gives the smallest residuals in the point source modelling.

\subsection{Simultaneous deconvolution}

Once the PSF is chosen, we proceed to a simultaneous deconvolution of the six individual images, in order to reach the best possible resolution and S/N ratio. As mentioned above, the MCS method is able to separate a pointlike source from a diffuse background, i.e., in our context, the QSO from its host galaxy.

We find that the best results are obtained in the following way:\\
(1) The position and intensity of the point source in each individual image are determined through deconvolution by ensuring that the image with the QSO point source subtracted contains a minimum amount of structure from the point source PSF (Airy rings, spikes,\ldots). In this way, we obtain individual host galaxy images.\\
(2) In a second phase, the MCS program is run on the six host galaxy images, which are simultaneously deconvolved, and a model is found which best represents this full set of observations.  The model is separated into two components: an analytical model accounting for the galaxy bulge, and a numerical background image accounting for all supplementary features (spiral arms, neighbors,\ldots).

The analytical model is chosen to follow a ``de Vaucouleurs" law with elliptical isophotes, i.e.\  an intensity profile:
\begin{equation}   
I(x)=I(x_1, x_2)=I_0 e^{-\big(Ax_1²+Bx_2²+Cx_1x_2\big)^{\frac{1}{8}}},
\end{equation}
where $x_1$ and $x_2$ are the two coordinates relative to the bulge center. The free parameters $I_0$, $A$, $B$ and $C$ are determined by the program. 

\begin{figure}
\includegraphics[width=16.5cm]{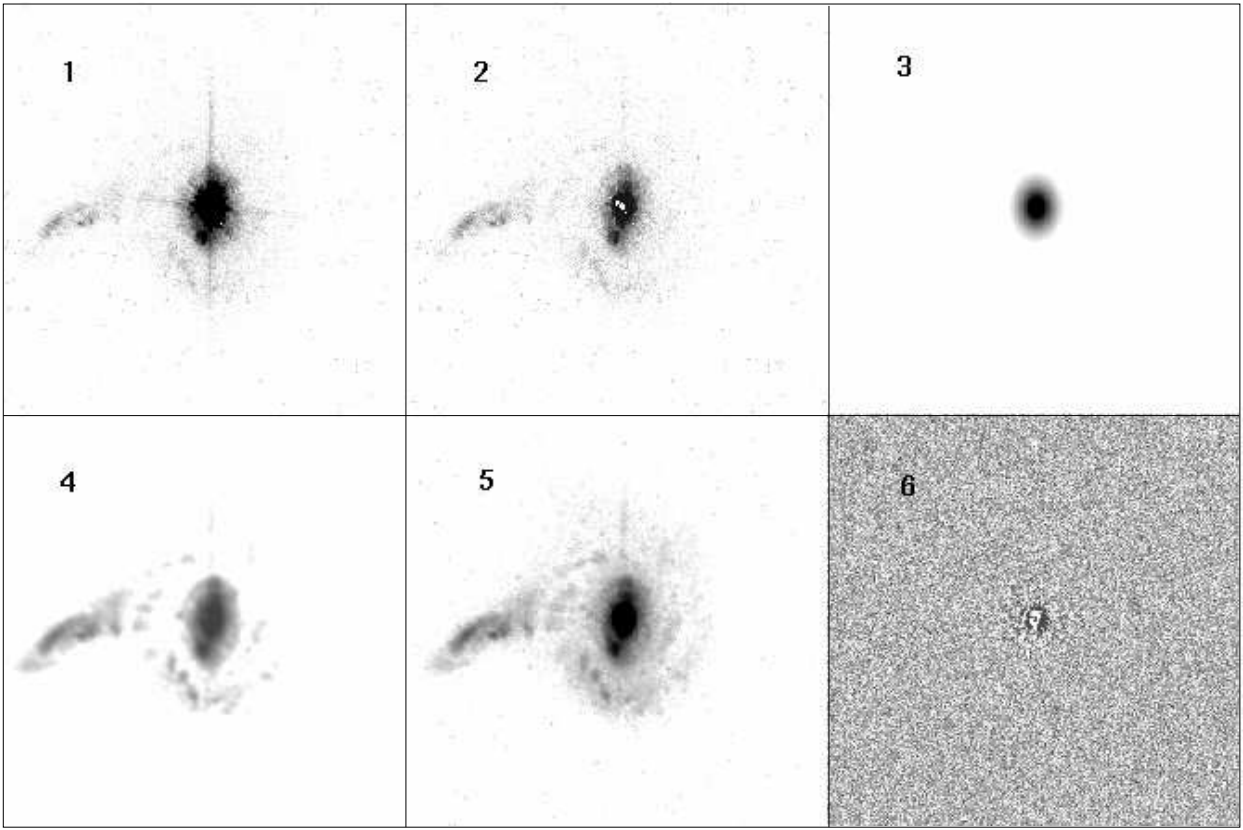}
\caption{Steps for QSO/host separation. 1) One of the long exposures of HE0354-5500 after reduction. 2) the point-like source has been removed from the observation. The PSF spikes are well removed. 3 to 6 are the results of the fit of the galaxy: 3) the analytical model found for the host galaxy bulge. 4) Numerical background. 5) The total fit of the galaxy (panel 3 + panel 4), the one used for analysis. Frame 6 displays the residuals (raw data minus total fit) of the whole process.}
\label{deconv}
\end{figure}

An example of the whole process is illustrated in Fig. \ref{deconv} for the QSO HE0354-5500. One sees that the residuals are very good, except around the very center. The reason for this misfit in the central region is that, in long exposures, both the PSF star and the QSO are saturated. As the weights of the pixels in the saturated regions have been set to zero, they do not constrain the fit. 

\placefigure{deconv}

\section{Spatially resolved spectra}
\begin{figure}
\centering
\includegraphics[width=12 cm]{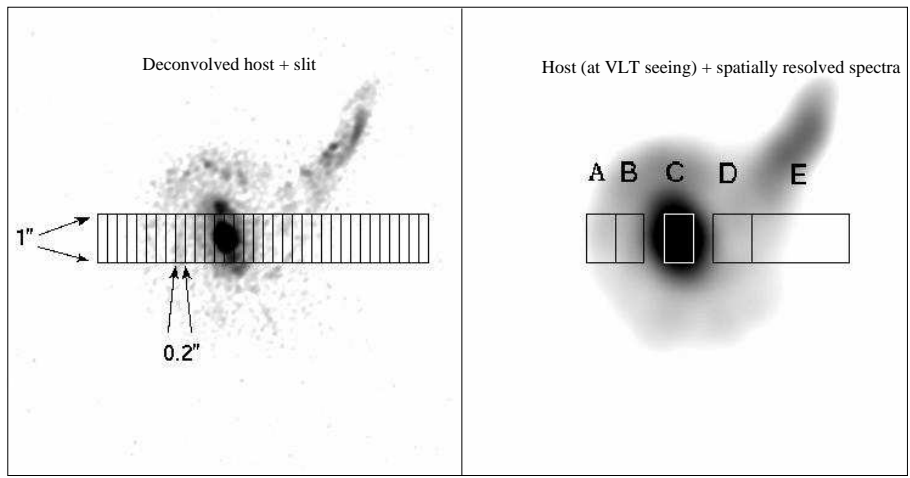}
\caption{Left: The VLT slit overlaid on the deconvolved image of the HE 0354--5500 host galaxy. Each column, corresponding to one pixel along the spatial direction of the slit, has a width of $0.2''$ and a height of $1''$. The deconvolved HST image is used to identify the pieces of spectra which correspond to specific parts of the host galaxy. Right: The spatially resolved spectra are superimposed on the deconvolved host galaxy, which is here reconvolved by a Moffat in order to reach the VLT spatial resolution.}
\label{0354VLT}
\end{figure}

\begin{figure}
\centering
\includegraphics[width=15 cm,height=15 cm]{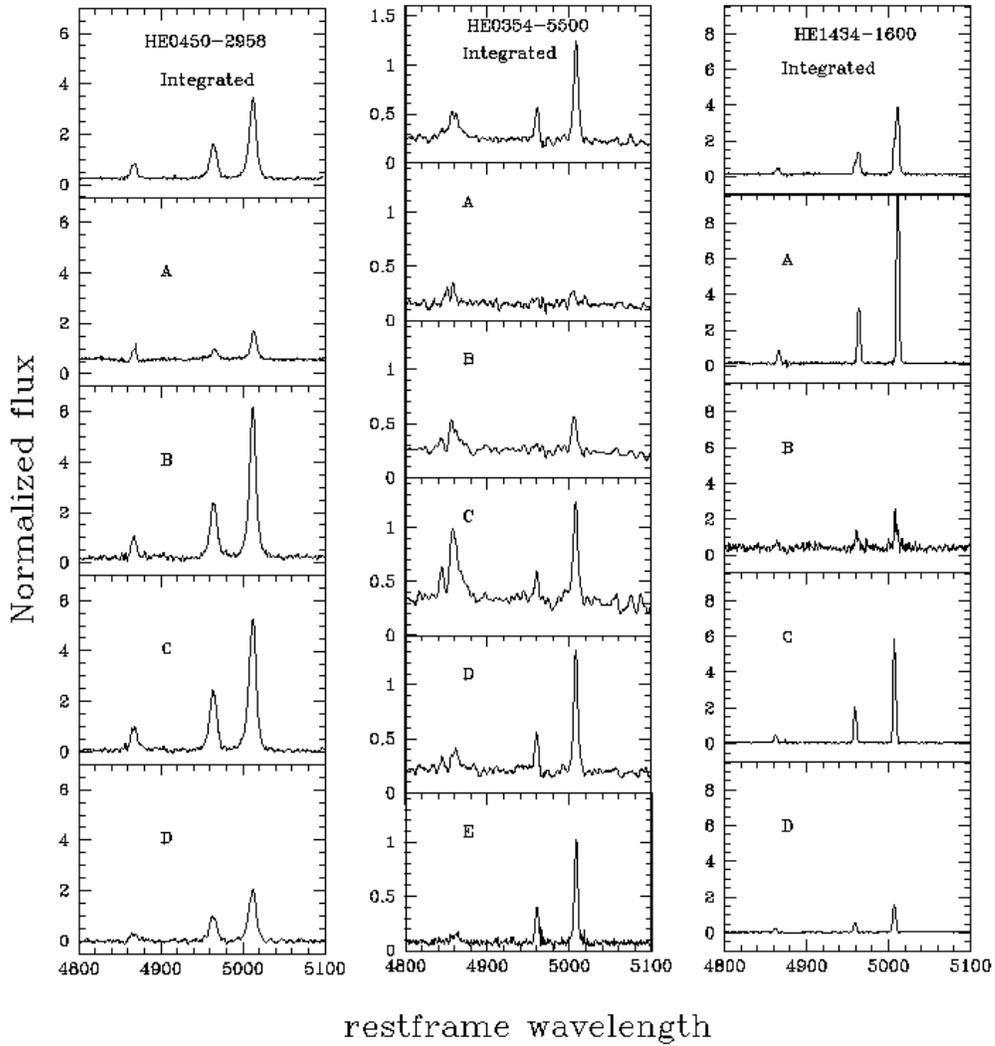}
\caption{Spatially resolved spectra of several objects around \hb-[OIII]. Each column scopes different regions of the QSO environment. These regions are labeled A-E and defined in Fig. \ref{1434_tout},\ref{0354_tout} and \ref{0450_tout}.} 
\label{regi_spec}
\end{figure}

By superimposing the projection of the VLT slit onto the HST host galaxy images, we can identify which parts of the galaxy contribute to each part of the spatially resolved spectra.  Of course, we have to take into account the seeing conditions during the ground-based spectroscopic observations.  This is done by convolving the host galaxy images with a 2D Moffat function derived from the 1D profile of the PSF stars observed simultaneously with the QSOs, and assuming circular symmetry of the PSF, which is accurate enough for the present purpose (see Fig.~\ref{0354VLT}). 
Even with the degraded resolution, it is still meaningful to separate the spectra into spatially resolved elements of different parts of the galaxy, such as spiral arms, bulge or halo. Several examples of extraction are given in Fig. \ref{regi_spec} for the \hb-[OIII] spectral region. These spectra are analyzed in Sect.\ref{results}.

\placefigure{0354VLT}
\placefigure{regi_spec}

\section{Spectral analysis}

Before any analysis, we correct the spatially resolved spectra for reddening, assuming a theoretical Balmer decrement \ha /\hb$=3.1$ if the region is ionized by the nucleus, and \ha /\hb$=2.85$ if it is ionized by stars,  using the Whitford reddening curves \citep{Miller}. The forthcoming detailed analysis focuses on the gaseous properties of the galaxies only, as the S/N for absorption lines is too weak.

\subsection{Diagnostic diagrams}

A diagnostic diagram uses line ratios to distinguish between the main ionization sources of the emitting gas, i.e. star formation, ionization by the QSO radiation or by shocks, as suggested by \citet{Baldwin} and \citet{Veilleux}. These diagrams were revised by \citet{Kewley}, who added a \textit{composite} region, that is, a region in the diagram where the ionization process is due to a mixture star formation and QSO radiation. 
The fluxes in the main emission features are measured from the spectra of the different galactic regions, as defined in the previous section.  We then compute the emission line ratios  [OIII](5007)/H$\beta$(4861), [NII](6583)$/$H$\alpha$(6563) and [OII](3727)$/$[OIII](5007) for each region and plot them in  diagnostic diagrams.  \\
Each region is differently affected by uncertainties in the measurements and so should have its own error bar. For clarity, when these uncertainties do not vary much from region to region, we only put one global error bar, which is representative of the different regions. Otherwise, specific error bars are displayed for each region.\\
Up to now, such diagnostic diagrams had only been used for Type 2 AGNs (i.e. narrow-line AGNs) and for the central regions of Type 1 (broad and narrow-line) AGNs. As far as we know, this is the first time that the hosts of Type 1 QSOs are analyzed with such diagnostic diagrams, except in \citet{Letawe} where the diagnostic diagrams are constructed from the integrated spectra.

\subsection{Radial velocity curves}

A very useful tool in the investigation of a QSO host is the radial velocity curve. We construct such a curve by comparing the wavelength shifts of some emission lines (e.g.,\  \ha\ or [OIII]) for every pixel along the slit. Then we transform the shifts in relative velocities with respect to the central QSO. An analysis of the velocity distribution of the host relative to the QSO is often fruitful to understand global tendancies in the dynamics and to model the mass distribution in the host. All the radial velocity curves and their analysis can be found in \citep{Letawe}. Some of them are discussed in Sect.\ref{results}.

\subsection{Line fits}

The \ha ($6563$\AA) line is partially blended with the neighboring [NII] lines (at $\lambda 6548$\AA\ and $\lambda 6583$\AA). In order to get accurate values for the line positions, widths and intensities, we model them as a sum of gaussian profiles.  

The two \nd lines have a theoretical intensity ratio of
\begin{equation}
\frac{\textrm{I}(\textrm{\nd}(6583))}{\textrm{I}(\textrm{\nd} (6548))}= 3,
\end{equation}
according to \citet{Oster}. This ratio is kept fixed in the fit as it only depends on the atomic structure  and not on the astrophysical environment. Moreover, both widths are set equal because the \nd  emissions arise from the same region, characterized by the same temperature and velocity distribution.

These emission lines are generally superimposed on a stellar spectrum, consisting in a continuum plus absorption lines.  In all the cases analyzed, the emission lines are so strong compared to the continuum that the latter can be locally approximated by a straight line, thus neglecting stellar absorptions and continuum curvature.

\begin{figure}
\centering
\includegraphics[height=7cm, width=8cm]{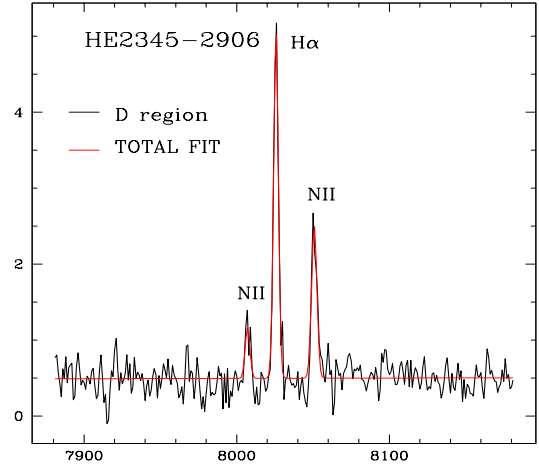}
\caption{Spectral modelling of the the \textit{D} region in HE $2345-2906$. The modeling of the spectrum by three gaussians and a linear continuum gives a very good fit.}
\label{HE2345}
\end{figure}

Taking these constraints and assumptions into account, we fit the spectrum around \ha\  using a sum of a linear continuum and three gaussians with two restrictions on the parameters: the intensity ratio between the \nd  lines as well as their widths. Some of the spectra are very well fitted by this method (see Fig. \ref{HE2345} for an illustration), but this fit turns out to be unsuccessful for some regions located far away from the nucleus in HE 0354--5500, HE 0450--2958, HE 1503+0228 and HE 0306--3301.

\placefigure{HE2345}
 
The three-gaussians model described above is unable to fit the data around the \ha\  line in those regions, and an additional broader component is necessary.  Adding a fourth gaussian generally allows to get an acceptable fit. 

\begin{figure}
\centering
\includegraphics[height=7cm, width=8cm]{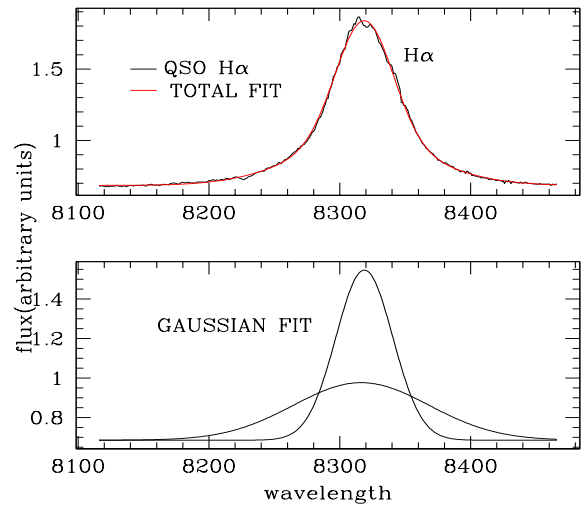}
\caption{The fit of the QSO \ha\ line of HE 0354--5500 by two gaussians.}
\label{QSO_0354}
\end{figure}

This fourth gaussian is much broader than the other components and its width is similar to that of \ha\  in the QSO spectrum.  We further test this similarity by assuming that the broad component has exactly the same profile as the QSO \ha\ line.  Under such an assumption, we  proceed to fit the spectrum in two steps:
\begin{itemize}
\item \textit{Step $1$}: we fit the \ha\ line of the central QSO spectrum by a sum of two gaussians (see Fig. \ref{QSO_0354});
\item \textit{Step $2$}: we fit the spectra of the host regions by a sum of $5$ gaussians: $3$ gaussians to account for the narrow \ha\ and \nd lines with the constraints described above. The two remaining gaussians are obtained from the fit of the QSO \ha\ line in the \textit{Step $1$}. Their parameters are constrained to have the same shape as the QSO \ha\ line: the intensity ratio, the shift between the centers and the widths are fixed. So, the only free parameters are the global intensity and the center the line.
\end{itemize}

\placefigure{QSO_0354}

However, we have to make sure that this broad line is not due to an imperfect subtraction of the QSO spectrum and that it indeed corresponds to a real feature.

The major source of error in deconvolution is the quality of the PSF. However, we have tested that changing the PSF within reasonable limits does not remove this broad component.
Moreover, if the broad line did arise from an imperfect subtraction of the QSO spectrum, it would most likely appear in all the regions of the host, because each wavelength is deconvolved independently in the spatial direction. But this broad component only appears in some regions, always located quite far from the nucleus.

In most instances, the broad component appearing in the host spectrum is redshifted in wavelength with respect to the QSO emission line, while an improper subtraction of the QSO spectrum would have most likely left a line at the same wavelength as in the central QSO.

We have also tested if the use of Lorentzian profiles could improve the fit to the data, as one might expect that the broader Lorentzian wings would weaken the need for an additional broad \ha\ component. However, it turns out that a combination of Lorentzians does not provide an adequate fit. Lorentzian profiles can be used to fit the spectra of some AGN when they have broader wings \citep{Veilleux91}, but in our case, the spectra are coming from different parts of the host galaxy and not from the AGN itself (the QSO spectrum has been removed).  In such a case, the emissions lines are expected to be gaussian-shaped and, indeed, gaussians are found to provide satisfactory fits in the majority of the cases (with the notable exception of those -- remote -- regions presenting a broad component).

\begin{figure}
\centering
\includegraphics[height=12cm,]{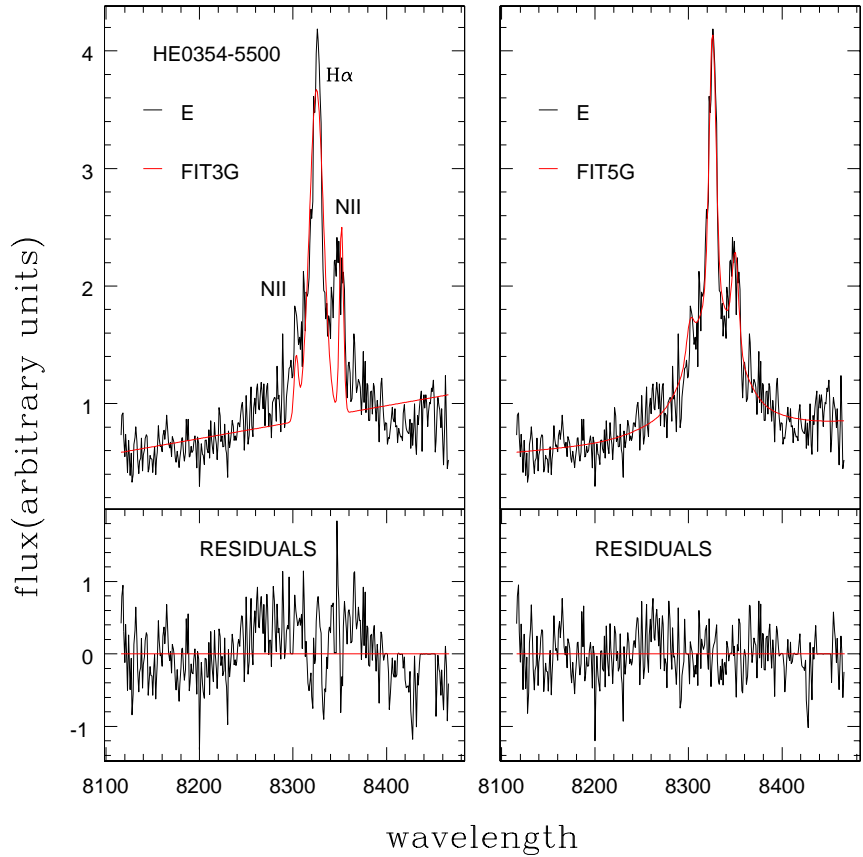}
\caption{Left: $3$-gaussian fit of the \textit{E} region around the \ha\ line in HE 0354--5500, along with its residuals. Right: $5$-gaussian fit, and residuals. Including a broad line with the QSO characteristics considerably improves the total fit of the \ha--[NII] region.}
\label{0354_fit}
\end{figure}

The improvement of the residuals of the fit when we add the QSO-like component in the case of HE 0354--5500, is shown in Fig. \ref{0354_fit}.

\placefigure{0354_fit}

\subsection{About the origin of the broad \textrm{\ha} line}

Which mechanism causes broad lines to be present in the spectrum of the host galaxy? The same kind of phenomenon is known in the radio galaxy \textit{Cygnus A} \citep{Ogle}, the Seyfert 2 galaxies NGC 1068 \citep{Simpson} and NGC 4388 \citep{Shields}, and in luminous infrared galaxies \citep{Tran}. In all those cases, the nucleus light is obscured or unobserved but emission coming from the nucleus is either reflected by dust or scattered by free electrons (for simplicity, we shall call it the reflected light). This reflection (or scattering) causes the light to be polarized. Spectro-polarimetry is then a good way to observe reflection of the nuclear emission in unresolved objects, and the analysis of the polarized reflected light allows to study indirectly some properties of the nucleus, such as its orientation or luminosity, while not being able to observe it directly because of dust obscuration \citep{Anton}.

A few cases of broad MgII or \hb\ line reflection, observed with direct spatially resolved spectroscopy, are reported in \citet{Dey} and \citet{Cid}, but, once more, only for Type 2 objects (for which the contrast host/QSO is much favourable than in Type 1 QSOs). In our case, only the \ha\ line is strong enough to be detected by reflection in some regions of the QSO environment.

In our sample of Type 1 QSOs, the similarity between the observed broad \ha\ lines in the host galaxy and the corresponding QSO \ha\ lines clearly suggests that reflection or scattering might be the explanation.
The main characteristics of the reflected broad lines are listed in Table \ref{broad} which gives:

\begin{enumerate}
\item The relative velocity shift of the reflected line with respect to the QSO emission line. A shift between the centers of the reflected \ha\ line and the QSO emission line clearly appears in some regions. In Table \ref{broad}, all those shifts have been transformed in relative velocity $\Delta \rm{v}$. Error bars on the measured shifts are estimated by shifting the center of the broad \ha\  until the fit becomes unacceptable. In a basic case of reflection, a shift between the emitted and the reflected light is interpreted in terms of motion of the reflecting medium. Thus, a shifted broad line reveals the presence of some material in motion with respect to the QSO. The fact that all shifts are positive shows that the reflecting medium is moving away from the QSO. The observed shift $\Delta \rm{v}$ is linked to the expansion velocity $\rm{v}_{exp}$ of the scattering particles in a resolved region of the host by the formula $$\Delta \rm{v}=\rm{v}_{exp}(1-cos\ \theta),$$ where $\theta$ is the angle between the direction of motion of the reflecting medium and the line of sight. 

Consequently, the measured shift must be seen as a lower limit to the expansion velocity of the reflecting medium with respect to the QSO.

\item The flux ratio $R=\dfrac{F_{reflected}}{F_{QSO}}$, between the flux of the reflected \ha\ broad line and the flux of the \ha\ line in the QSO spectrum.

\item The equivalent width (EW) of the reflected line compared to the QSO emission line. If we assume that the medium reflects both the \ha\  line and the underlying continuum, and that a stellar continuum may also be present, the reflecting regions must have an EW equal (if there is no stellar continuum) or smaller (in absolute value) than the QSO emission line.  This is indeed the case.

\item The ratio $\frac{H_{\alpha}}{H_{\beta}}$, evaluated on the galactic narrow component of \ha\ for the region containing the broad line, reveals the presence of dust, as reddening increases this ratio above the nominal value of 3.1 or 2.85, depending on the excitation mechanism.

\item The deduced electron temperature and an upper limit on this temperature. If the scattering is due to electrons, the reflected light must be broadened according to the random motion of the scattering electrons. In practice, we convolve the QSO \ha\ line by a gaussian representing the electron velocity distribution  (corresponding to a given  temperature) and use this new profile instead of the QSO \ha\ line in the $5$-gaussians fit. The best fit gives an estimate of the electron temperature $T_{e^-}$. Increasing the electron temperature until the fit becomes unacceptable sets an upper limit $T_{max}$.
\end{enumerate}

\begin{table*}
\scriptsize
\centering
\caption[]{Characteristics of the reflected brad \ha\ line in the spectrum of the host galaxies.}
\begin{tabular}{lcccccc}
\hline
Object &Region&Rel. vel. &Flux ratio $R$ & reflected \ha/QSO \ha & $\frac{\textrm{\ha}}{H_{\beta}}$ &$T_{e^-}-T_{max} $($10^5K$) \\ 
     &        &  ($\kms$)       &       &$\vert EW \vert$  & \\
\hline\hline
\vspace{2pt}
HE 1503+0228&A&$+200^{+120}_{-280}$&$3.09\pm 0.09\ 10^{-3}$&$71/345$&$5.38$&$2.27-6$ \\ 
\vspace{2pt}
HE 1503+0228&E&$+200^{+240}_{-280}$&$3.36\pm 0.10\ 10^{-3}$&$98/345$&$4.28$&$5.55-12.8$\\
\vspace{2pt}
HE 0306--3301&F'&$+620^{+150}_{-150}$&$1.03\pm 0.05\ 10^{-3}$&$280/365$&-&$0.73-1.88$ \\
\vspace{2pt}
HE 0354--5500&E&$+458\pm 180$&$4.06\pm 0.12 \ 10^{-3} $&$117/116$&$3.66$&$1.02-4$\\
\vspace{2pt}
HE 0450--2958&B&$+745^{+250}_{-280}$ & $9.33\pm 0.20 \ 10^{-3}$&$422/558$&$3.49$&$0-1.35$\\
\hline
\end{tabular}
\label{broad}
\end{table*}

All regions containing a reflected line are faint in the HST images (except in HE 1503+0228), revealing a weak (if any) stellar continuum. Indeed, one might imagine thet the reflected \ha\ line may be present in many regions, but only observable when the stellar continuum is weak enough, that is, in low luminosity regions of the images. 

It is hard to determine whether the phenomenon at work is mostly reflection by dust or scattering by free electrons. First, the $\frac{H_{\alpha}}{H_{\beta}}$ ratios in Table \ref{broad} indicate that all regions showing a broad \ha\ line contain some dust. Second, all fits with zero broadening can be considered acceptable, which shows that dust reflection is a viable hypothesis.  On the other hand, in all objects except HE0450--2958, the best fit is obtained with a line having a non-zero broadening, which points towards electron scattering.  Moreover, the temperatures obtained under the assumption of electron scattering appear reasonable and consistent with the estimates of \citet{Dey} ($T_{e^-}\approx 3.10^5K$). Finally, all the spectra of these regions show high ionization lines, which indicates that free electrons are present. It is thus very difficult to decide between the two different alternatives. Both may contribute to various degrees in different objects.

\section{Results on individual objects}
\label{results}

The main characteristics of the host galaxies\footnote{The bulge ellipticity is defined in terms of our analytical model by $\epsilon = \frac{B-A}{(B+A)cos(2\theta)}$, where $\theta=0.5 \arctan(\frac{C}{B-A}).$}, as derived from the HST images, are summarized in Table \ref{resu}.  The individual cases are discussed below.  We start by the least powerful nuclei, HE 1503+0228 and HE 2345--2906, and proceed in order of increasing nuclear luminosity.

\begin{table*}
\centering 
\caption[]{Global properties of the host galaxies. Morphology, nucleus to host luminosity ratio (N/H), bulge ellipticity (bulge ell.) and orientation of the main axis counted positive East of North ($\theta$ in degrees). HE0450--2958 is not listed, as no host was detected.}
\begin{tabular}{lcccc}
\hline
Name & Morphology & N/H & bulge ell. & $\theta$   \\
     \hline\hline
HE 1503+0228 & Spiral  & $1.9361$ &  $0.0116$ & $10$ \\
HE 2345--2906 & Spiral & $0.9039$ & $0.0294$ & $-43$ \\
HE 1434--1600 & Ellipt. & $1.7833$ & $0.1698$ & $-29$\\
HE 0306--3301 & Spiral & $7.4079$ & $0.3921$ & $37$ \\
HE 0354--5500 & Spiral & $2.1568$ & $0.2534$ & $-101$ \\
\hline
\end{tabular}
\label{resu}
\end{table*}

\subsection{HE 1503+0228}

The analysis of this object has already been presented in \citet{Courbin}.  The HST images confirm their conclusions, but bring significant additional information.
 
\begin{figure}
\centering
\includegraphics[width=16.5cm]{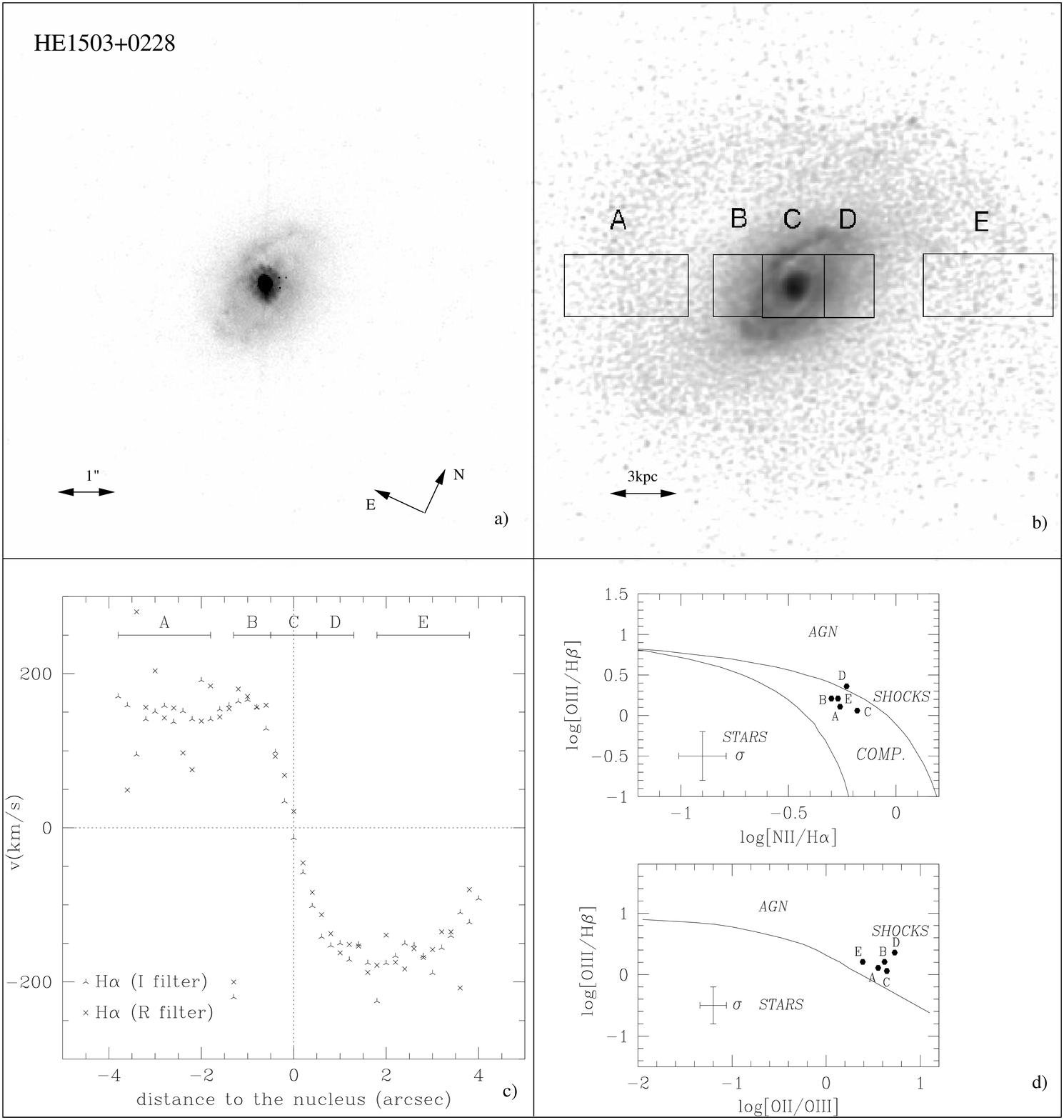}
\caption{HE 1503+0208. a) Reduced image. b) Deconvolved host galaxy with the different slits used to extract the spatially resolved spectra in overlay. The image reveals a regular spiral host surrounded by a kind of halo, which could be wider spiral arms, as suggested by the spectral analysis. c) Radial velocity curve. d) Diagnostic diagrams.}
\label{1503_tout}
\end{figure}

At first sight, the host of HE 1503+0228 seems to be a quite mundane spiral galaxy (Fig. \ref{1503_tout}). However, the deconvolved image reveals a faint, extended component with approximately circular symmetry, surrounding the whole system at $\approx 6.5$ kpc from the center.

\begin{figure}
\centering
\includegraphics[width=9cm]{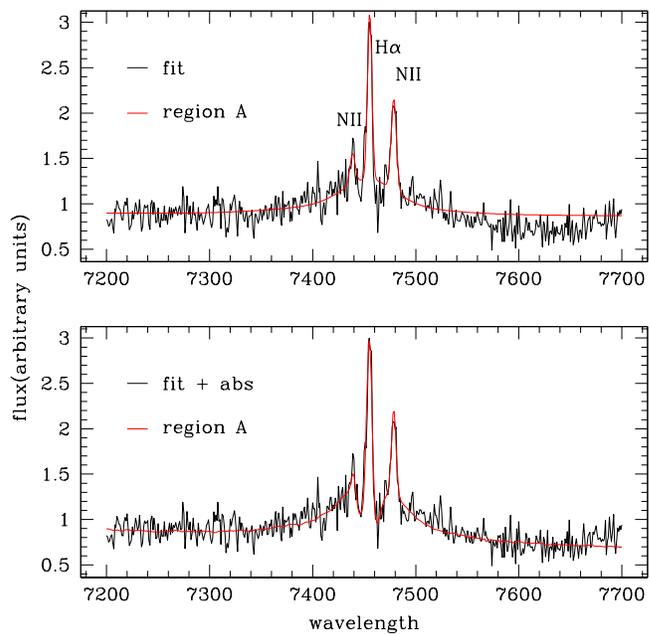}
\caption{HE 1503+0228: The $5$-gaussians fit of the outer region (top image) is improved when a stellar absorption component is added (bottom image).}
\label{1503_fit}
\end{figure}

As can be seen from Fig. \ref{1503_tout}, the different regions are clustered in the same area of the diagnostic diagram, indicating a mix of ionization by shocks and by stars. However, this faint extended component (regions \textit{A} and \textit{E}) differs from the bulge and from the inner arms in that it shows a broad \ha\  component superimposed on the narrow emission lines and on a significant continuum.  Among all the broad reflected \ha lines, those detected in HE 1503+0228 have the smallest EWs.  This suggests that the contribution of the stellar continuum to the total spectrum is significant.  This conclusion is further supported by the fact that the spectrum around \ha\ cannot be properly fitted with a $5$-gaussian model.  Taking into account a stellar \ha\  in absorption allows to improve the fit (see Fig. \ref{1503_fit}).  The $\chi^2$ minimization leads to an EW of the absorption line of $11.5$ \AA\ for region \textit{A} and $16.5$ \AA\ for region \textit{E}, which is larger than generally found in integrated galactic spectra (see, e.g., \citet{Kenni}, \citet{Tress}).  However, the S/N of our spectrum is rather low in regions A and E, so that the uncertainties are quite large (the one sigma uncertainty on the intensity of the absorption is about $35\%$). The large EW of the absorption line suggests that the stellar spectrum in these external regions is dominated by a young stellar population.  This is consistent with the conclusion drawn from the diagnostic diagram, which shows a mix of ionization by shocks and by stellar light, and with the large Balmer decrement measured in these regions (Table \ref{broad}).  Indeed, if significant stellar \ha\  is present in absorption, it should also be seen in \hb. Such a stellar absorption should introduce a larger underestimate of the relatively weak \hb\  narrow emission line, when compared to the much stronger \ha\  emission (thus explaining the high Balmer decrement measured in these regions).

\placefigure{1503_fit}

The presence of a broad reflected line in the outer regions \textit{A-E} and not in the inner ones \textit{B-D} raises the following problem.  Since these regions show otherwise similar spectra, with about the same ionization level, one could wonder how the QSO light can cross the inner spiral arms without being absorbed or reflected, to reach the outer regions, and be reflected there.  The most natural explanation seems to be that these outer reflecting regions are not located in the galactic plane.  This is supported by the deconvolved HST image, which shows that, while the inner region, containing the spiral arms, has a significant ellipticity, the outer part is much more circular.  If it was located in the same plane, and had a circular symmetry, the inclination angle of the galactic plane should introduce the same apparent ellipticity as for the inner region.  Its circular morphology suggests that it is either in a different plane, close to the plane of the sky, or that it has a roughly spherical symmetry (e.g.\  a halo populated with young stars).  These peculiarities are supported by the radial velocity curve (Fig.\ref{1503_tout}), which shows a rather sharp drop in the outer \textit{E} region (the spectrum is more noisy in the \textit{A} region).  This drop could be interpreted as due either to a rotation in a plane closer to the plane of the sky or to a geometry closer to spherical.\\
In the case of non active galaxies, outer rings surrounding spirals are supposed to be created either by gravitational interactions (merger, collision, or even tidal interactions by a neighboring galaxy), or by resonance effects with a bar, if any \citep{Buta1,Buta2,Buta3}. In our case, the peculiar geometry and spectra of HE 1503+0228's outer ring suggest that some gravitational interaction probably happened in the past 
(e.g.\   a minor merger or a tidal interaction with a passer-by).  This past interaction might also be related to the ignition of the QSO activity.

\subsection{HE 2345--2906}

\begin{figure}
\centering
\includegraphics[width=16.5cm]{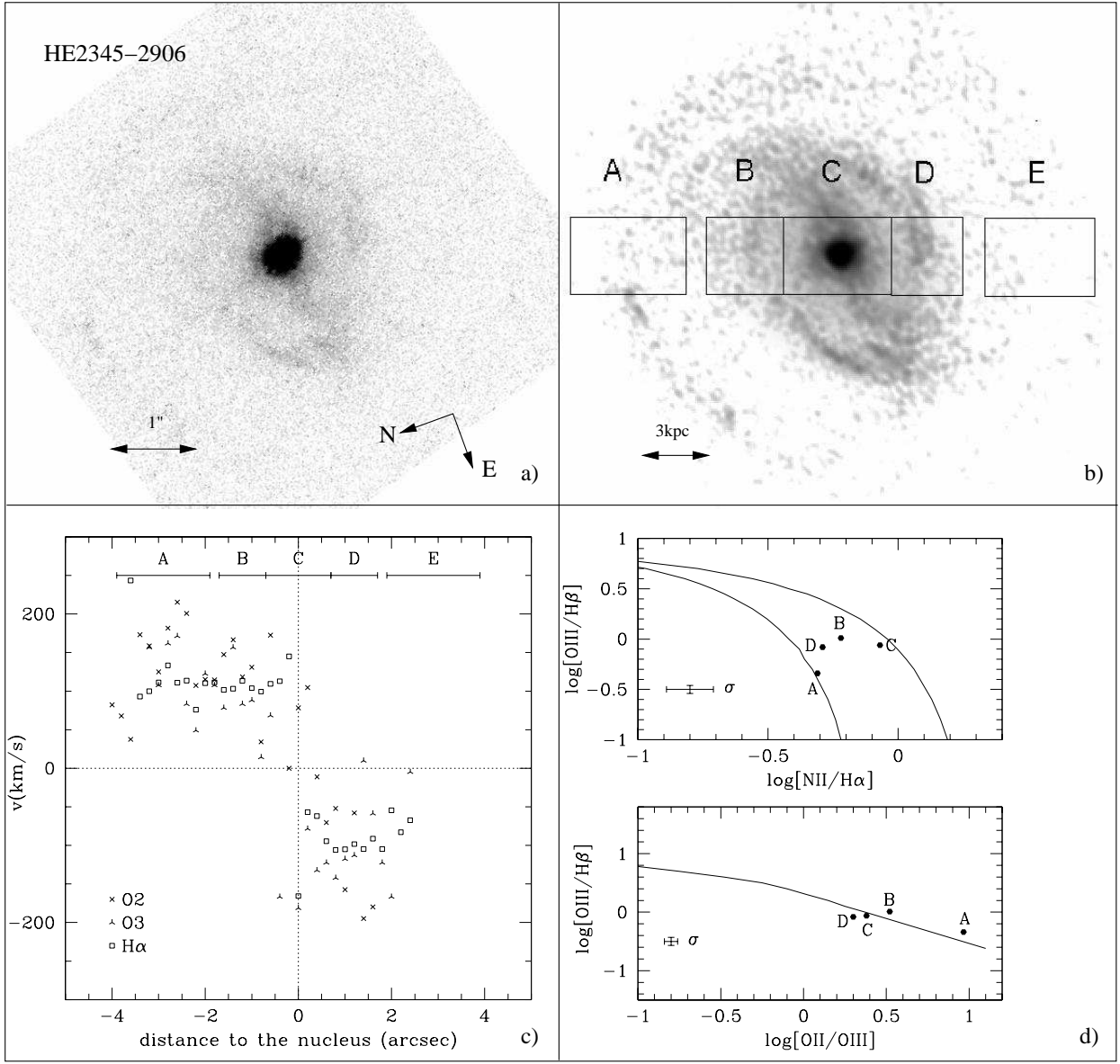}
\caption{HE2345--2906. a) Reduced image. b) Deconvolved host galaxy with the slits in overlay. The host is clearly a barred spiral galaxy. Wider arms can be found $\approx 8$kpc from the center. c) Radial velocity curve. d) Diagnostic diagrams.}
\label{2345_tout}
\end{figure}

As seen from Fig. \ref{2345_tout}, the host is a barred spiral, with a prominent 8 kpc-long bar. Two spiral arms originate from the extremities of the bar and surround the close-to-spherical bulge. A wider and fainter arm, located 8 kpc from the center, surrounds the whole system. The diagnostic diagrams show that the emission regions lie in the \textit{composite} region of the diagram, which means that the gas ionization is probably due to a mix of shocks, stellar light and QSO radiation.

A bar in spiral host galaxies is often considered as an efficient means of funnelling matter into the central regions and feeding the AGN (e.g. \citet{Sellwood1999,Crenshaw2003}).  As shown in Fig. \ref{HE2345}, no broad reflected component is detected.  Moreover, all the spatially resolved spectra are typical of a classical spiral galaxy.

\subsection{HE 1434--1600}

The spectra of this object were analyzed in \citet{Letawe3}, and their results are confirmed by the present study.

\begin{figure}[h]
\centering
\includegraphics[width=16.5cm]{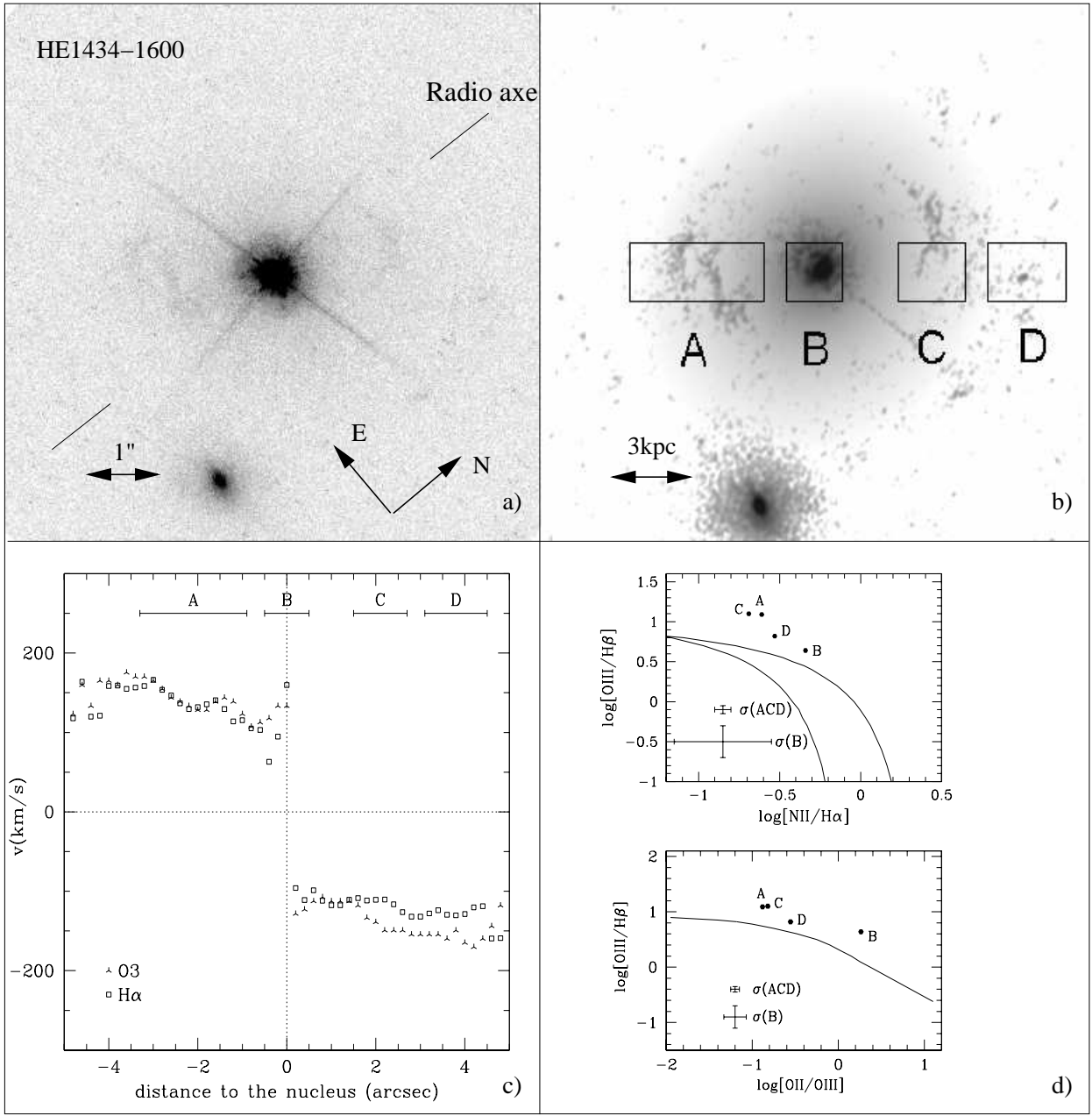}
\caption{HE1434--1600. a) Reduced image with the axe of the two radio lobes \citep{Condon} superimposed. b) Deconvolved host galaxy with the slits in overlay, revealing filaments on both side of the center, and an undisturbed neighboring elliptical galaxy. c) The radial velocity curve tells us about the motion of the gaseous filaments. d) Diagnostic diagrams.}
\label{1434_tout}
\end{figure}

The deconvolved image (Fig. \ref{1434_tout}) shows that, apart from the central QSO, the system is made of at least three components: (1) an elliptical host completely taken into account by our analytical model (whose central part corresponds to region \textit{B}), (2) a neighboring elliptical galaxy, and (3) some filamentary structures (regions \textit{A}, \textit{C} and \textit{D}) on both sides of the center and $\sim 3$ to $8$ kpc far away from it. 
The ionization diagrams of Fig. \ref{1434_tout} show that the filaments consist of gas highly ionized by the QSO.  The central region \textit{B} lies closer to the transition between ionization by shocks and by the AGN. 

The analysis of the radial velocity curve (Fig. 15 of \citet{Letawe3}), enlightened by the ACS image,  shows that the filaments are moving with respect to the central QSO, at a speed of at least 150 km/s (which is the measured radial velocity), and slightly increasing outwards. They most probably consist of gas blown away and accelerated by the QSO radiation.  

The presence of extended regions of highly ionized gas in an elliptical galaxy is unusual. It might have been brought there by a past interaction (maybe related to the QSO ignition), and its presence is revealed to us only thanks to the QSO strong ionization field. The bipolar structure of the gas is probably related to an orientation effect of the QSO. Radio observations of HE 1434-1600 \citep{Condon} reveal two lobes extending up to $980$kpc, North and South of the QSO, which is roughly the same orientation as the gaseous filaments (Fig. \ref{1434_tout}).  Thus, the ionized gas might only be a part of the whole gas present the host galaxy. 

\subsection{HE 0306--3301}

\begin{figure}
\centering
\includegraphics[width=16.5cm]{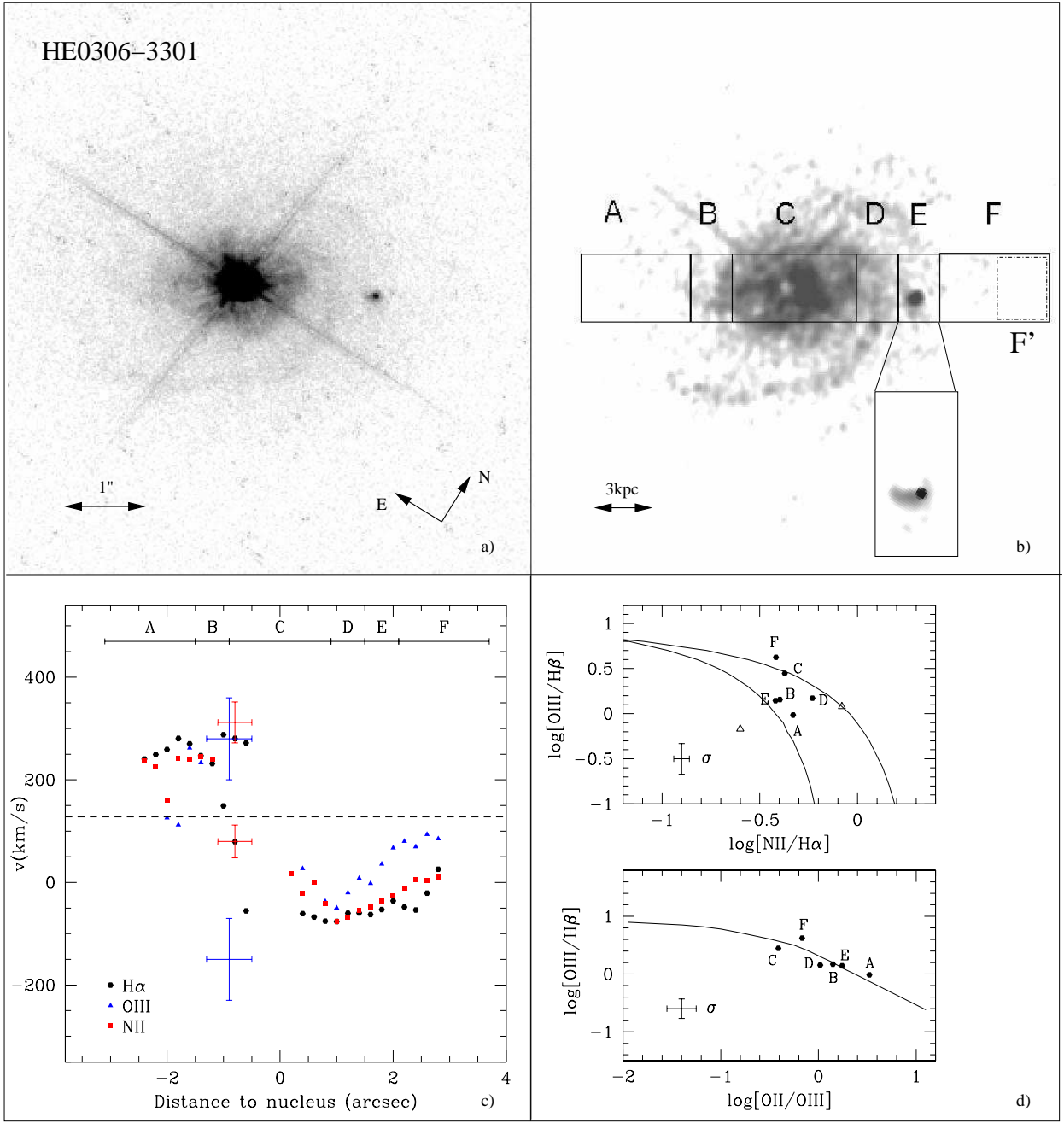}
\caption{HE 0306--3301. a) Reduced image. b) Deconvolved host galaxy with the slits in overlay. As different smoothings had to be applied to the central regions and to the spot, a close-up of the spot with the appropriate smoothing is shown (with a point source).  c) Radial velocity curve. In the region around $-1\arcsec$, the emission lines are resolved into 2 components.  For \ha, the velocities of the 2 components are indicated.  For the other emission lines, they are summarized by the crosses whose height corresponds to the range spanned by the measured velocities. The noise in the curve and the few discrepant points are due to the weakness of the emission lines. d) Diagnostic diagrams. The different ionizations measured on the separated components in the B-C region are indicated by open triangles.}
\label{0306_tout}
\end{figure}

Figure \ref{0306_tout} shows the deconvolved image of the host galaxy of HE $0306-3301$. The deconvolution of this particular case has been harder to handle as the QSO is highly saturated and the central region affected by the loss of linearity extends to a larger area. This explains the residual spikes and apparent gaps around the very center.

At first glance, the host looks like a rather mundane spiral galaxy.  However, deeper scrutiny reveals that this first impression is far from telling the whole story.

First, the spiral arms look asymmetric and are more prominent in the N-W part of the host.  In this region, a compact source (that we shall call `the spot') is found in (or close to) one of the spiral arms, at about 5 kpc from the QSO.  A close-up on that region is shown in Fig. \ref{0306_tout}.  It is well fitted by a point source, plus a faint extension in the direction of the QSO.  In the V-band, the spot is about 1000 times fainter than the central QSO.

The diagnostic diagrams presented in Fig. \ref{0306_tout} show that all regions contain gas whose ionization level is in the transition region between stellar and AGN (or shocks).  There is a tendency for the ionization to increase from the S-E to the N-W, i.e.\  when crossing the galaxy in the direction of the spot and further away.

The VLT spectra are not deep enough to allow extracting the spectrum of the spot.  However, they reveal a very unusual radial velocity curve over the whole galaxy (Fig. \ref{0306_tout}).  The N-E side of the host has a rather constant velocity of $\simeq 250\kms$ relative to the central QSO. However, when approaching the QSO position, in the inner 1 arcsec, the emission lines become resolved into two components.  We have fitted two gaussians to separate these two components and find that, in all three main emission lines, i.e. \ha, [NII] and [OIII], the highest velocity component remains at about the same velocity as in the outer regions.  Its position in the diagnostic diagrams shows a mix of ionization sources, as found in the other regions of the galaxy.  The lowest velocity component displays radial velocities approaching zero in the QSO rest frame, as well as an ionization degree typical of star-forming regions.

The depth of the spectra does not allow to build a radial velocity curve based on the absorption lines.  However, an average absorption lines radial velocity is determined on each side of the QSO by correlating the host spectrum with a template spiral galaxy spectrum in the spectral region between the [OII] and \hb\ lines, which is devoid of measurable emission lines.  We obtain $+175 \kms$ for the S-E side and $+80 \kms$ on the N-W side, towards the spot. The error bars are of the order of $50 \kms$.  This means that the average velocity of the bulge stars is shifted by $\sim 130 \kms$ with respect to the QSO.  

Velocity shifts between different QSO emission lines are common, QSO lines being generally blueshifted with respect to their host galaxy restframe.  One can thus wonder how to obtain a reliable measurement of the QSO redshift.  This question was investigated by \citet{Letawe}, who compared the velocity shifts of different QSO emission lines with the host galaxy radial velocity, in five cases of spiral hosts with regular morphology and symmetrical rotation curves.  In such cases, the host redshift can be determined accurately and without ambiguity.  They found that, in these symmetrical cases, the velocity measured from the tip of the QSO \ha\ and \hb\ emission lines agrees with the host velocity, while the bulk of these lines indicates a blueshift (related to the well-known extended blue wing).  Narrow QSO emission lines are also generally blueshifted.

Following \citet{Letawe}, we have used the tip of the \ha\ and \hb\ emission lines to estimate the QSO redshift.  When compared to the results of \citet{Letawe}, which show an average velocity shift compatible with zero and a scatter of about $20 \kms$, the $130 \kms$ difference found in this case is clearly an outlier.  The velocity difference we measure between the QSO and the bulge of its host galaxy can be considered secure at the $6 \sigma$ level.

\begin{figure}
\centering
\includegraphics[width=9cm]{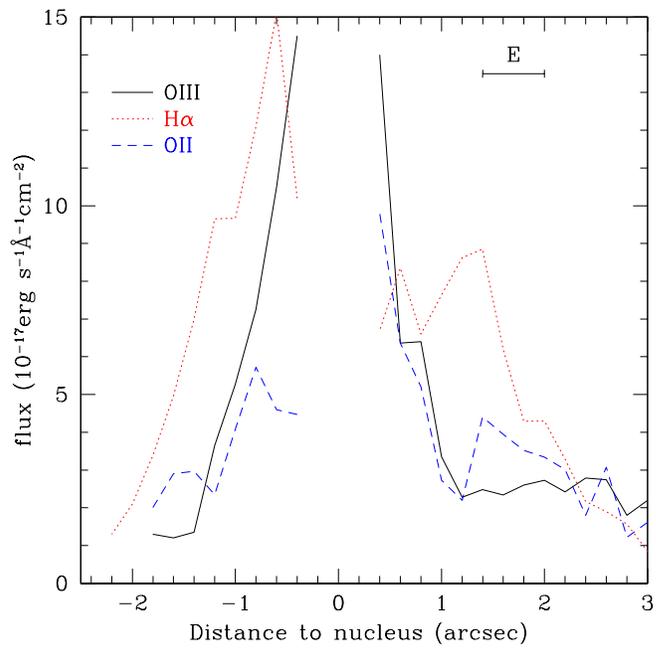}
\caption{HE 0306--3301: measured flux in the host galaxy spectrum for different emission lines ([OIII], [OII] and \ha) as a function of distance from the QSO. The region \textit{E} corresponding to the spot is indicated.}
\label{flux}
\end{figure}

Considering now the radial velocity curve on the N-W side of the QSO, Fig. \ref{0306_tout} shows that all three emission lines indicate roughly the same velocity in the inner 1 arcsec, $\sim -50 \kms$ with respect to the QSO, or $\sim -180 \kms$ with respect to the average velocity of the stars.  Then, the radial velocities of the three emission lines start to increase at different rates. The [OIII] emission line increases first, followed by [NII] and then by \ha.  The differences are significant and maximal at $\sim 100 \kms$, in the region of the spot.  Figure \ref{flux} shows the flux in these three emission lines as a function of distance from the QSO.  It shows that, in the region where the radial velocities disagree, the flux distribution is very different in the three lines. \ha\ is maximal and [OIII] minimal at about $1.5´´$ from the QSO, i.e.\  close to the spot.  [OII] has intermediate strength.  This is typical of a star formation region.  Note that the absolute magnitude of the spot,
 as determined from the HST image, amounts to $M_V \sim -17.3$, which is also rather typical of starburst galaxies \citep{Allen}.  The spot might thus correspond to the core of an intense star-forming region.

Moving further away from the QSO, the radial velocity curves tend to converge to roughly the same value, $v \sim +120 \kms$ with respect to the QSO.  The narrow emission lines become quite weak in this region which, in its outermost part (labelled \textit{F'} in Fig.\ref{0306_tout}), also displays a broad (reflected) \ha\ line.  Combining the radial velocity shifts measured on the narrow lines and on the reflected broad line indicates that the reflecting medium is moving away from the QSO at a speed  $v\sim 370 \kms$, and at an angle $\theta \sim 70\degr$ with respect to the line-of-sight.

We are far from fully understanding the origin of all the peculiarities found in this QSO host galaxy.  The discrepant velocities measured from different narrow emission lines may be due to the fact that several gas clouds with different ionization stages contribute to the spectral emission in some regions of the host.  In other regions, their velocities are sufficiently different for the different components to be resolved.  Moreover, the strange behaviour of the radial velocity curve in the region of the spot, as well as the velocity shift between the QSO and the galactic bulge, indicate that this system is undergoing violent phenomena, which might be related to a collision between two galaxies whose images are superimposed on the plane of the sky.

\placefigure{flux}

\subsection{HE 0354--5500}

The HST image (Fig. \ref{deconv}) reveals a distorted host galaxy, with the probable presence of a few spiral arms, as well as an extended structure to the North, suggestive of a (nearly edge-on?) galaxy in violent interaction with the host.

\begin{figure}
\centering
\includegraphics[width=16.5cm]{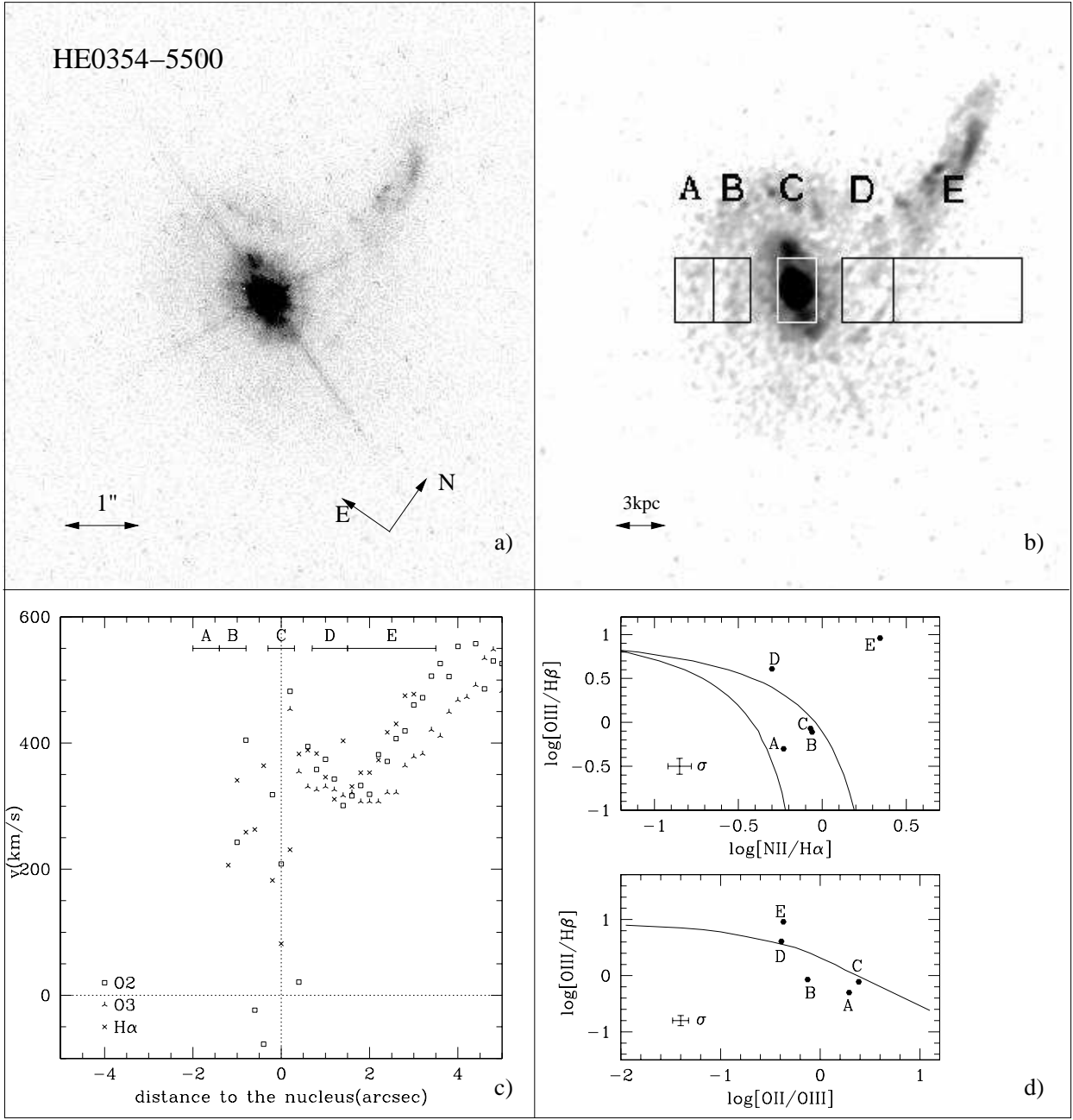}
\caption{HE0354--5500. a) Reduced image. b) Deconvolved image with the slits in overlay (a zoom on the bulge is presented in Fig.\ref{centre}). c) Radial velocity curve. d) Diagnostic diagrams.}
\label{0354_tout}
\end{figure}

The diagnostic diagrams  (Fig. \ref{0354_tout}) show that the source of ionization changes as one moves along the slit, from the S-E (region \textit{A}) to the N-W (region \textit{E}) with a more important contribution from stellar light on the S-E parts (opposite to the colliding galaxy), a mixture of sources in the central regions and ionization completely dominated by the AGN in the N-W region \textit{E}.

Figure \ref{0354VLT} shows that region \textit{E} may contain some contribution from the two colliding galaxies.  However, its spectrum displays only emission lines, with no detectable continuum.  Moreover, these emission lines extend to the outer regions, where no flux is measured on the HST image.  We thus conclude that the spectrum is completely dominated by emission from AGN-ionized gas.  The radial velocity curve shows velocities increasing from $+300$ to $+550 \kms$, as one moves away from the center.

Moreover, the spectrum of this region \textit{E} displays a clear broad component in the \ha\ emission (Fig.  \ref{0354_fit}). This may be interpreted as either reflection of the QSO light by dust, or scattering by free electrons.  The relative velocity of this broad component with respect to the QSO is $v=+460 \kms$, which is close to the average radial velocity measured on the narrow emission lines.  This near equality shows that the gas moves away from the central QSO, in a direction opposite to us, and at an angle $\sim 45\degr$ with respect to the line-of-sight. 

Combining these two pieces of informations leads to the plausible interpretation that region \textit{E} contains gas blown away by the AGN radiation.  The radiation pressure, acting continuously as the gas moves away from the center, produces a continuous acceleration and a speed increasing from about $400$ to $600 \kms$ in the region where we can measure it (i.e. outside the visible image of the host galaxy).

This expanding cloud of ionized gas also scatters the QSO light.  As it displays a range of velocities, the electron temperature, measured from the broadening of the \ha\  line (Table \ref{broad}), must be considered as an upper limit. However, this range of velocities ($\sim 200\kms$) is too small to sufficiently broaden the line, as $\sim 2000\kms$ would be required.

\begin{figure}
\centering
\includegraphics[width=12cm]{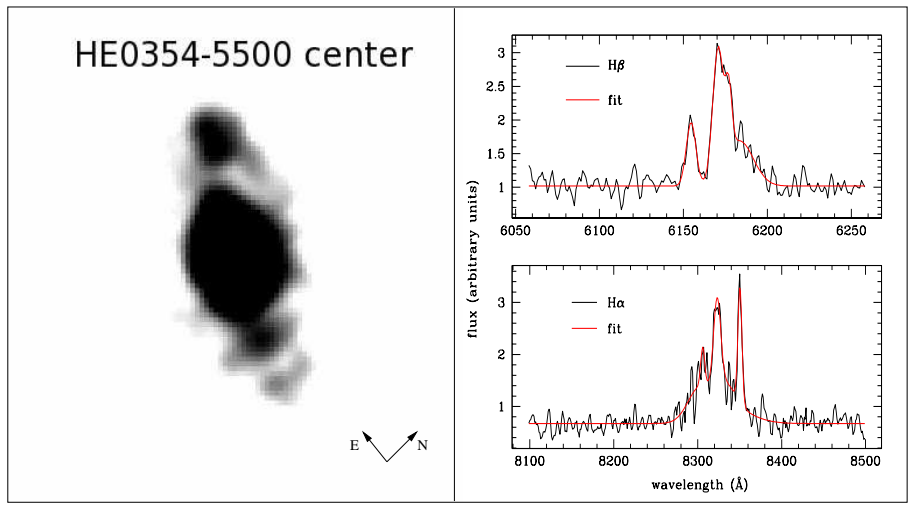}
\caption{Left: deconvolved central region of HE 0354-5500 revealing different bright areas. Right: fit of the spectrum of the central region \textit{C} (see Fig.\ref{0354_tout}) around \hb\ (top panel) and \ha (bottom panel).}
\label{centre}
\end{figure}

Other interesting features appear in the bulge region.  The H$\beta$ line displays a complex structure (Fig. \ref{centre}, top right), which can be interpreted as the sum of several components moving at different speeds.  Indeed, the deconvolved HST image (Fig. \ref{centre}, left) shows, next to the bulge, three bright regions which, because of the spread of light due to the seeing, cannot be separated in the VLT spectra.  Figure \ref{centre} shows that the \hb\ line can be fitted by 4 gaussians, which is also the number of emission regions seen on the HST images.

Due to the presence of the [NII] lines, the H$\alpha$ profile is not easy to interpret.  However, Fig. \ref{centre} shows that a model with the same components as detected in H$\beta$ can provide a satisfactory fit of the observations.

\placefigure{centre}

On the other hand, only the strongest component is detected in the [OIII] lines.  As this component is at the same radial velocity as the QSO, it probably corresponds to the bulge. Its position in the diagnostic diagrams (region \textit{C}) shows a mixture of ionization sources, with a dominant contribution by shocks.  On the other hand, the other components, with no detectable [OIII], should be dominated by stellar ionization.  They probably correspond to starbursts initiated by the galactic collision, which might also be the cause of the QSO ignition.

\subsection{HE 0450--2958}

\begin{figure}
\centering
\includegraphics[width=16.5cm]{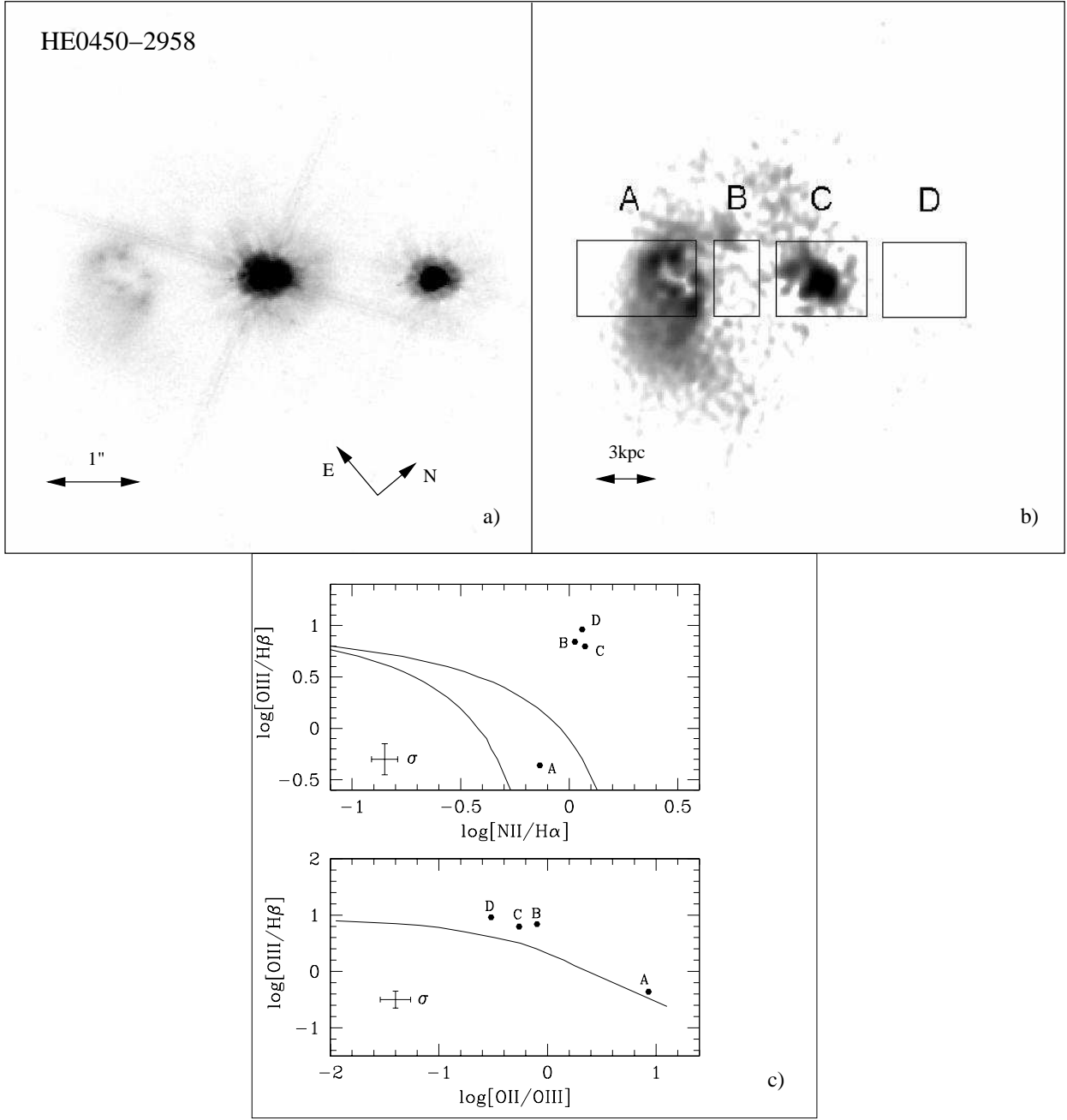}
\caption{HE 0450--2958. a) Reduced image. b) Deconvolved image with the nucleus and the star removed thanks to the deconvolution process. c) Diagnostic diagrams.}
\label{0450_tout}
\end{figure}

This very special case (Fig. \ref{0450_tout}) has been analyzed in \citet{Mag2}. No host is detected on the HST images, which allow to set an upper limit implying that it is underluminous by a factor of at least 6 for a QSO of that absolute magnitude. \citet{Merritt}, noting that the QSO spectrum displays rather narrow broad emission lines, suggest that the black hole might be of rather low mass, but with a high accretion rate.

A companion galaxy is found at about 7 kpc S-E of the QSO.  It has a strongly distorted shape, probably resulting from a collision.  Careful removal of the QSO light reveals a compact emission region (\textit{C}) right next to the QSO.  The size of this object is $\sim 1$kpc and its spectrum shows that it consists of gas ionized by the QSO, with no detectable stellar continuum.

The diagnostic diagrams in Fig. \ref{0450_tout} (bottom) show the clear separation between the companion galaxy spectrum and the other regions. The companion galaxy ionization is due to a mixture of star formation and shocks, whereas the other regions appear to be ionized by the QSO itself. This result tends to favour the view that, if there is star formation near the QSO, it is much weaker than the contribution of the ionization by the AGN.

\begin{figure}
\centering
\includegraphics[width=12cm]{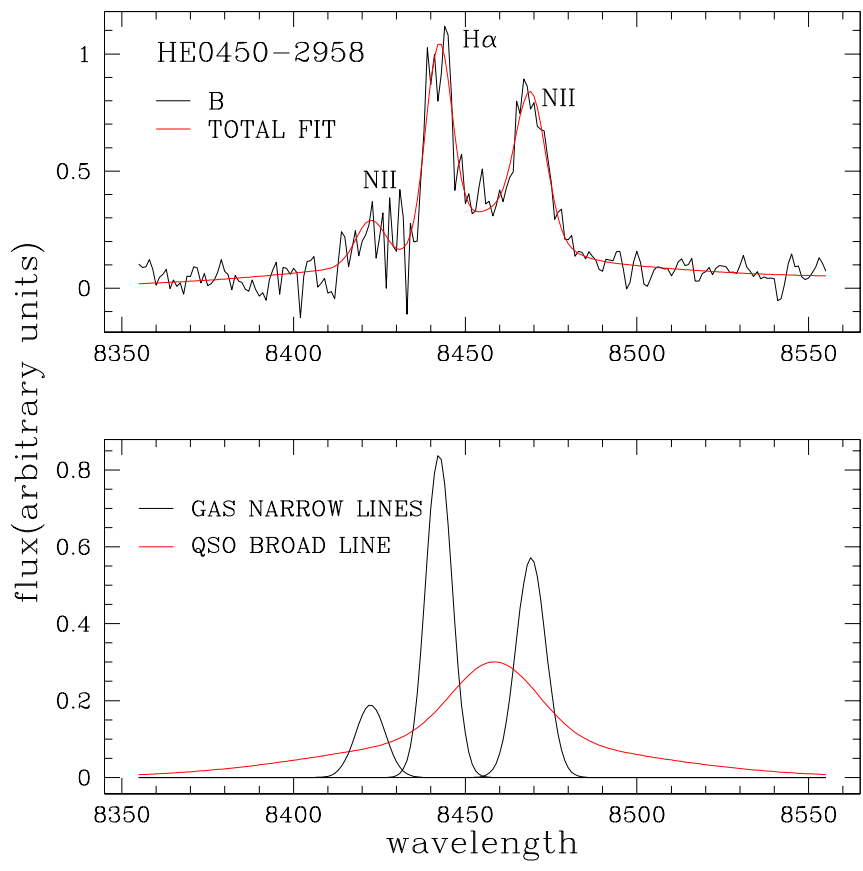}
\caption{Top image: fit of the region \textit{
B} in HE 0450--2958 around the \ha\ line. Bottom image: the different lines fit. The broad \ha\ line is strongly shifted compared to the narrow line.}
\label{HE0450_fit}
\end{figure}

A broad \ha\ component is detected in the {\it B} region (Fig. \ref{HE0450_fit}), in between the QSO and the companion galaxy. Its shift relative to the QSO \ha\ line center corresponds to a relative velocity of $+745\kms$, the highest one measured in the present study. It has no relation either with the velocities measured in the companion galaxy, which range from $-60$ to $+200\kms$ relative to the QSO.  It is also completely different from the radial velocity measured from the narrow emission lines ($+160\kms$).  Assuming that the matter responsible for the narrow line emissions and for the reflection (or scattering) of the QSO light share the same motion (it may indeed belong to the same gas cloud), we can combine the two measured radial velocities to derive a speed of $\sim +900\kms$ and an angle of $\theta \sim 80\degr$.  Thus, this gas moves away from the QSO and towards the companion galaxy, in a direction close to the plane of the sky and at a speed close to $1000\kms$.  This is one more indication that the system is undergoing violent interactions and that this powerful QSO has a profound influence on its large-scale environment.
\placefigure{HE0450_fit}

\section{General conclusions}

The present study has shown that: 
\begin{enumerate}
\item A joint analysis of spatially resolved VLT spectra and high-resolution HST images, both carefully processed with the MCS method, is a powerful way to characterize the QSO host galaxies; it allows to discover phenomena which would not be accessible through imaging or spectroscopy alone.
\item The \ha\ line, in regions far away from the center, has a broad component due to reflection of the nuclear light by dust or scattering by free electrons. Our study provides an extension of this phenomenon, only observed for Type 2 QSOs (i.e.\  objects in which the central engine is obscured by dust), to Type 1 QSOs. Evidence for extended scattered light has been independently found by \citet{Borguet}.
\item In most cases, the AGN has long-range effects on its environment.  This important point in the AGN-starbust interaction, already known for both Type 1 (with the ``proximity effects", see \citet{Bajltik}, \citet{Ferrarese,Sijaki}) and Type 2 QSOs \citep{Kauff}, is confirmed here in a completely different and independent way for Type 1 QSOs.
\end{enumerate}

We now summarize the main results on the individual objects, in order of increasing AGN luminosity.

The two lower luminosity objects, HE 1503+0228 and HE 2345--2906, which lie close to the somewhat arbitrary limit between Seyfert galaxies and QSOs, are spirals.  One of these spiral hosts has a very prominent bar and the other one has an extended arm or shell of gas and stars, probably reminiscent of a past interaction.

HE 1434--1600 consists in a massive elliptical in gravitational interaction with a rather modest neighbor.  Moreover, it contains shells of ionized gas blown up by the central QSO.

HE 0306--3301, which has about the same luminosity as the previous one, looks like a spiral but has an extremely odd radial velocity curve.  The central QSO has a radial velocity of $\simeq 130$km/s with respect to the galactic bulge.  Moreover, a compact spot is found at $\sim 6$ kpc from the center.  This spot, which might correspond to an intense starburst region, seems to add further perturbations to the radial velocity curve.

HE 0354--5500 looks like a (barred ?) spiral undergoing a violent collision.  It shows compact starburst regions close to the center and an extended gas component most probably accelerated by the QSO radiation, up to distances as large as 8 kpc.

Finally, the most luminous object, HE 0450--2958 (which is a powerful infrared emittor and may also be the youngest QSO in the sample) has no detected host galaxy, but has a strongly distorted galactic companion and has long range influence on its neighborhood.

Although our sample is quite small, the fact that the most powerful QSOs seem associated to the most dynamically perturbed host galaxies might not be a mere coincidence and we might be tempted to conclude that dynamical perturbations are efficient means of feeding the central AGN. Of course, even if the idea of a link between AGN activity and interactions is already 20 years old \citep{Sanders}, larger samples of bright radio-quiet QSOs with combined high resolution imaging and 3D spectroscopy are needed to further check these findings, to shed more light on the dynamical processes responsible for the ignition and fueling of the most active AGNs and to allow a more precise assessment of their impact on their galactic-scale environment.

\acknowledgments

This work was supported by PRODEX Experiment Agreement 90195 (ESA and PPS Science Policy, Belgium).

\onecolumn

\end{document}